\documentclass[a4paper,dvips,12pt]{article}
\usepackage{epsfig}
\usepackage{amssymb}
\usepackage{cite}
\newcommand{\bea}{\begin{eqnarray}}   
\newcommand{\eea}{\end{eqnarray}}   
   
\newcommand{\NPB}[3]{\emph{ Nucl.~Phys.} \textbf{B#1} (#2) #3}   
\newcommand{\PLB}[3]{\emph{ Phys.~Lett.} \textbf{B#1} (#2) #3}   
\newcommand{\PRD}[3]{\emph{ Phys.~Rev.} \textbf{D#1} (#2) #3}

\def\simlt{\stackrel{<}{{}_\sim}}

\title{   
\vspace*{-0.8cm}   
\begin{flushright}   
\normalsize{      
IEM-FT-213/01\\
IFT-UAM/CSIC-01-12\\   
\texttt{hep-ph/0104112}}\\ 
\end{flushright}    
\vspace{1cm}
\Large\textbf{One-loop Higgs mass 
finiteness in supersymmetric Kaluza-Klein theories~\footnote{Work 
supported in part by CICYT, Spain, under contract AEN98-0816,
and by EU under contracts HPRN-CT-2000-00152 and HPRN-CT-2000-00148.}}
\vspace*{.5cm}
\author{\large
{\bf A.~Delgado, G.~v.~Gersdorff, P.~John and  M.~Quir{\'o}s}\\ \\
\emph{Instituto de Estructura de la Materia (CSIC), Serrano 123,}\\
\emph{E-28006 Madrid, Spain.}}}
\date{}   
\begin{document}
\maketitle
\thispagestyle{empty}
\vspace*{.5cm}
\begin{abstract}
We analyze the one-loop ultraviolet sensitivity of the Higgs mass in a
five-dimensional supersymmetric theory compactified on the orbifold
$S^1/\mathbb{Z}_2$, with superpotential localized on a fixed-point
brane. Four-dimensional supersymmetry is broken by Scherk-Schwarz
boundary conditions. Kaluza-Klein
interactions are regularized by means of a brane Gaussian distribution
along the extra dimension with length
$l_s\simeq\Lambda^{-1}_s$, where $\Lambda_s$ is the cutoff of the
five-dimensional theory. The coupling of the $n$-mode, 
with mass $M^{(n)}$, acquires the $n$-dependent factor 
$\exp\left\{-(M^{(n)}/\Lambda_s)^2/2\right\}$, which makes it to decouple for
$M^{(n)}\gg \Lambda_s$.
The sensitivity of the Higgs mass on $\Lambda_s$ is strongly
suppressed and quadratic divergences cancel by supersymmetry. The one-loop
correction to the Higgs mass is finite and equals, for large values of
$\Lambda_s$, the value obtained by the so-called KK-regularization.

\end{abstract}
\vspace{4.cm}   
   
\begin{flushleft}   
April 2001 \\   
\end{flushleft}

\newpage
One fundamental problem in particle physics is to understand
the origin of electroweak symmetry breaking (EWSB) that
leads to the pattern of vector boson and fermion masses in the
Standard Model (SM).  The only known perturbative mechanism (the Higgs
mechanism) requires a fundamental scalar, the Higgs boson, which
acquires a vacuum expectation value (VEV) and therefore breaks
spontaneously the electroweak symmetry. In the SM, the radiative
corrections to the squared Higgs mass are quadratically sensitive
to the cutoff (which jeopardizes the consistency of the theory), while
in its minimal supersymmetric extension (MMSM), the sensitivity is
only logarithmic. 

It is a common belief that in supersymmetric  
theories with one extra dimension radiative corrections to scalar 
masses are not sensitive (at least at one-loop) 
to the ultraviolet (UV) cutoff of the
theory~\cite{antoniadis,delgado,barbieri,hall,quiros}. A similar result
holds in (four-dimensional) theories at finite temperature, where the size of
the extra dimension is given by the inverse temperature. Present 
calculations have been performed summing over all Kaluza-Klein (KK)-modes.  
This is known as KK-regularization and ignores that a five dimensional (5D)
theory must be seen as an effective theory below the cutoff
($\Lambda_s$). This fact has recently been accounted by
imposing the sharp cutoff $\Lambda_s$ on the momentum integration and 
truncating the summation in the KK-tower~\cite{nilles} to modes
with $M^{(n)}<\Lambda_s$. This leads to quadratic
divergences because the sharp truncation of the KK-modes spoils the tower
structure of the 5D theory.

In this letter we analyze a regularization where KK-modes are not
truncated, but instead the brane is extended over the extra dimension with a
finite length $l_s\simeq1/\Lambda_s$ by a Gaussian distribution. 
This regularization is suggested from string theories~\cite{strings}.
In particular we will study the UV sensitivity of the
Higgs mass in the prototype model presented in Ref.~\cite{quiros},
although the results are much more general and should also 
apply to other mass calculations and models.
The model is based on a 5D $N=1$ theory whose massless modes constitute
the usual four dimensional (4D) $N=1$ MSSM. Supersymmetry breaking is a
bulk phenomenon induced by Scherk-Schwarz (SS) boundary 
conditions~\cite{SS1,SS2} and radiative
breaking is triggered by the presence of a bulk top/stop hypermultiplet.

The setup of the model is as follows. The 5D space-time is
compactified on $\mathcal{M}_4 \times S^1/\mathbb{Z}_2$. The orbifold
$\mathcal{M}_4\times S^1/\mathbb{Z}_2$ has two fixed points at
$y=0,\ell$ where the two 3-branes are located ($\ell\equiv \pi R$ is
the length of the segment). There are two types of fields: those living
in the bulk of the extra dimensions, similar to the untwisted states
in the heterotic string language ($U$-states), and those living on the
branes localized at the fixed points, similar to the heterotic string
twisted states ($T$-states). We will assume that gauge fields are
$U$-states organized in $N=2$ gauge multiplets 
$\mathbb{V}=(V_\mu,\lambda_1;\Sigma+i V_5, \lambda_2)$. Even and odd components
under $\mathbb{Z}_2$ are separated by a semicolon,
as $(even;odd)$. Notice that only
the even fields have zero modes. Matter
fields can either be $U$- or $T$-states. In order to have Yukawa
interactions to generate (after EWSB) fermion masses only two
possibilities for the localized superpotential are
allowed~\cite{quiros}: they are of the form $UUU$ or $UTT$. We will
deal only with the latter~\footnote{The case with $UUU$ was recently
studied in Refs.~\cite{barbieri,hall}.}, as the former one is
suppressed by a relative factor $(\ell \Lambda_s)^{-1}$. So we will consider
the model where all $SU(2)_L$ doublets, $H_{1,2}$, $Q$, and $L$,
are localized in the boundary $y=0$, and the singlets
$\mathbb{U}=(U,\Psi_U;U^{\prime}, \Psi_{U^{\prime}})$,
$\mathbb{D}=(D,\Psi_D;D^{\prime}, \Psi_{D^{\prime}})$ and
$\mathbb{E}=(E,\Psi_E;E^{\prime}, \Psi_{E^{\prime}})$ are in the
bulk. Both, even and odd components, form independent chiral
multiplets.

Supersymmetry breaking is done by compactification using the
SS-mechanism~\footnote{We are using the $N=2$ $SU(2)_R$ global symmetry 
(or, more specifically, its $U(1)_R$ subgroup 
preserved upon orbifold action) as
the generator of supersymmetry breaking. Thus, only fields
transforming under $SU(2)_R$ will get a mass.}, with parameter
$\omega$ and mass spectrum as follows (for details, see
Ref.~\cite{quiros}):

\begin{itemize}
\item
KK-modes of gauge bosons and right-handed matter fermions 
have masses $M^{(n)}=n \pi/\ell$. There are massless states 
corresponding to $n=0$.

\item
KK modes of gauginos and scalar partners of right-handed 
matter fermions get masses
$M^{(n)}=(n+\omega) \pi/\ell$. Zero modes acquire soft masses $\omega/R$.

\item
Left-handed fermions and sfermions, and the Higgs sector remain massless
at tree-level.
\end{itemize}

We will introduce a superpotential with Yukawa couplings that, along
with the gauge couplings, will induce radiative masses to 
brane scalars. We will regularize the interactions
with localized states by assuming that the brane has a finite
extension of length $l_s\sim 1/\Lambda_s$ along the fifth dimension
with a Gaussian distribution
\begin{equation}
\label{gaussian}
f_G(y;l_s)=\frac{1}{\sqrt{2\pi}l_s}e^{ -\frac{y^2}{2 l^2_s}}.
\end{equation}
The rapid fall-off of the Gaussian will produce an exponential
suppression of the coupling of KK-modes with masses $M^{(n)}\gtrsim
\Lambda_s$ which leads to an effective cutoff of these modes without
spoiling the tower structure of the 5D theory. Notice that the so-called
KK-regularization corresponds to the distribution (\ref{gaussian}) in the
limit $l_s\to 0$, according to the limit 
$\delta(y)= \displaystyle\lim_{l_s\to 0} f_G(y;l_s)$.

We then use the superpotential
\begin{equation}
\label{superpotential}
W=\left[h_U\, Q\,H_2\,U+h_D\, Q\,H_1\,D+h_E\, L\,H_1\,E
\right] f_G(y;l_s),
\end{equation}
where we denote with the same symbols both the supermultiplets and their
scalar components.
The use of the Gaussian distribution in (\ref{superpotential}) will
change the couplings between the Higgs and matter fields, with respect
to the common formalism where a $\delta$-distribution is used
(KK-regularization). We will
calculate these coupling using the off-shell
formalism~\cite{peskin1}. 

The Lagrangian for the fields involving
$h_U$-couplings is:
\begin{eqnarray}
\label{l5}
&&\mathcal{L}_Y=\int dy\Biggl\{|F_U|^2+f_G(y;l_s)\Biggl[|F_Q|^2+|F_{H_2}|^2+
\nonumber\\
&&\phantom{1^{1^1}}\left.\phantom{1^{1^1}}
\left(\frac{\partial W}
{\partial U}(F_U-\partial_y U^{\prime})
+\frac{\partial W}{\partial Q}F_Q+\frac{\partial W}{\partial H_2}F_{H_2}+
 \frac{\partial^2 W}
{\partial Q \partial U}\Psi_Q\Psi_U+h.c.
\right)
\Biggr]
\right\},
\end{eqnarray}
where $F_U$ ($F_Q$) is the $F$-component of the $U$ ($Q$) superfield,
$\Psi_U$ ($\Psi_Q$) its fermionic component, and $U^{\prime}$ 
is the odd scalar of the $\mathbb{U}$
hypermultiplet, which couples to the brane through its derivative with
respect to the extra dimension.

Taking into account the non-trivial twist of the bosonic fields due to
the SS boundary conditions, the Fourier expansion of the fields is given by
\begin{eqnarray}
\label{fourier}
F_U&=&\sum_n \cos\left[\frac{(n+\omega) \pi y}{\ell} \right]
F^{(n)}_U \nonumber\\
U&=&\sum_n \cos\left[\frac{(n+\omega) \pi y}{\ell} \right]U^{(n)}\nonumber\\
U^{\prime}&=&\sum_n \sin\left[\frac{(n+\omega) \pi y}{\ell} \right]
U^{(n)} \nonumber\\
\Psi_U&=&\sum_n \cos\left[\frac{n \pi y}{\ell}\right] \Psi^{(n)}_U.
\end{eqnarray}
Using the identity~\footnote{Notice that, strictly speaking, the integral
in (\ref{inte}) should be performed over the orbifold length. However since the
gaussian distribution decays exponentially fast we are allowed to extend the
interval of integration over the whole real axis. This provides a good 
enough approximation and is not changing our finite final result.}
\begin{equation}
\label{inte}
\int_{-\infty}^{\infty} dy \cos\left[\frac{ (n+\omega)\pi}{\ell}y\right] 
f_G(y;l_s)= e^{-\frac{(n+\omega)^2\pi^2}{2(\ell
\Lambda_s)^2}}
\end{equation}
and integrating out the auxiliary fields we end up with the following 
Lagrangian for Yukawa interactions:
\begin{eqnarray}
\label{l4}
\mathcal{L}_Y=
&&\sum_{n=-\infty}^{\infty} \left[h_t^2\,e^{-\frac{(n+\omega)^2\pi^2}
{(\ell \Lambda_s)^2}}
\left\{ |H_2|^2 |U^{(n)}|^2+|H_2|^2 |Q|^2+|Q|^2 |U^{(n)}|^2
 \phantom{1^{1^{1^{1^1}}}}\right.\right.\nonumber\\
&&\left.\phantom{1^{1^1}}\left.-\left
(\frac{(n+\omega)\pi}{\ell} U^{(n)} H_2 Q+h.c.\right)\right\}
+h_t\,e^{-\frac{n^2\pi^2}{2( \ell \Lambda_s)^2}} 
H_2 \Psi_Q \Psi^{(n)}_U +h.c. \right].
\end{eqnarray}

Notice that the Lagrangian (\ref{l4}) can be interpreted as one where the
couplings of heavy KK-modes $U^{(n)}$ are model dependent and suppressed as:
\begin{equation}
h_t^{(n)}=h_t \exp\left\{ -\frac{1}{2}\left(\frac{M^{(n)}}{\Lambda_s}
\right)^2\right\}.\
\label{acoplos}
\end{equation}
In this way the decoupling of heavy KK-modes occurs without spoiling the
tower structure of the theory.

We can now calculate the contribution at one-loop to the
Higgs mass as \cite{delgado}: 
\begin{equation}
\label{masa}
m_{H_2}^2=\Delta m^2(\omega)-\Delta m^2(0),
\end{equation}
where
\begin{equation}
\Delta m^2 (\omega)= 2 h_{t}^2 N_c
\ell^{\,2}\sum_{n=-\infty}^{\infty}\int \frac{d^4p
}{(2\pi)^4}\frac{e^{-\frac{(n+\omega)^2\pi^2}
{(\ell\Lambda_s)^2}}}{(\ell p)^2+
(n+\omega)^2\pi^2}\,\,\,,
\label{sum}
\end{equation}
$N_c$ being the number of colours.

Using the Schwinger representation for the propagators and performing the $p$ 
integration over $0\leq |p|\leq \Lambda_s$ one gets for $\Delta m^2(\omega)$
the expression
\begin{eqnarray}
\Delta m^2(\omega)= \frac{h^2_t N_c}{8\pi^2\ell^{\,2}}\sum_{n=-\infty}^{\infty}
\int_{0}^{\infty}ds\,\,e^{-(n+\omega)^2\pi^2[s+\frac{1}
{(\ell\Lambda_s)^2}]}
\frac{1}{s^2}\left[1-e^{ -s\ell^{\,2}\Lambda_s^2}(1+s\ell^{\,2}\Lambda_s^2
)\right].
\end{eqnarray}
Notice that the apparent divergence of the integrand at $s\to 0$ (UV limit)
cancels because of the presence of the cutoff $\Lambda_s$. 

The integral
over $s$ can then be performed and yields
\begin{equation}
\Delta m^2(\omega)= 
\frac{h^2_t N_c}{8\pi^2\ell^{\,2}}\!\!\sum_{n=-\infty}^{\infty}
\left[(\ell\Lambda_s)^2
-(n+\omega)^2\pi^2\log\frac{(n+\omega)^2\pi^2+(\ell\Lambda_s)^2}
{(n+\omega)^2\pi^2}
\right]
e^{-\frac{(n+\omega)^2\pi^2}{(\ell\Lambda_s)^2}}.
\label{twoterms}
\end{equation}
This contribution to the Higgs mass contains a quadratically divergent
term. However, this quadratic divergence is canceled by supersymmetry.
In fact, using the Poisson resummation formula
\begin{equation}
\label{poisson}
\sum_{n=-\infty}^{\infty} g(n+\omega)=\sum_{n=-\infty}^{\infty} 
e^{-2\pi in\omega}\int_{-\infty}^\infty dz~e^{-2\pi inz}g(z)
\end{equation}
which, for the case of the Gaussian simply gives 
\begin{equation}
\sum_{n=-\infty}^{\infty} e^{-\frac{\pi^2}{(\ell \Lambda_s)^2}(n+\omega)^2}=
\frac{\ell\Lambda_s}{\sqrt{\pi}}\sum_{n=-\infty}^{\infty} e^{-2\pi i n \omega}
e^{-n^2(\ell \Lambda_s)^2},
\end{equation}
we find for the first term of (\ref{twoterms})
\begin{equation}
\frac{h^2_t  N_c}{8\pi^{5/2}\ell^{\,2}}(\ell\Lambda_s)^3\sum_{n=-\infty}^{\infty}
e^{-2\pi i n \omega} e^{-(\ell\Lambda_s)^2n^2}.
\label{first}
\end{equation}
The corresponding contribution to $m^2_{H_2}$ can then be written
(after including supersymmetric terms) as
\begin{equation}
\frac{h^2_t  N_c}{8\pi^{5/2}\ell^{\,2}}
(\ell\Lambda_s)^3\sum_{n=-\infty}^{\infty}
\left(e^{-2\pi i n \omega}-1\right) e^{-(\ell\Lambda_s)^2n^2},
\label{two}
\end{equation}
whose $n=0$ term cancels from supersymmetry and whose leading
contribution in the limit $\ell\Lambda_s\to \infty$ is provided by the 
$n=1$ mode. This one behaves like
$(\ell\Lambda_s)^3e^{-(\ell\Lambda_s)^2}$ and clearly vanishes when
$\ell\Lambda_s\to \infty$.

\begin{figure}[ht]
\centering
\epsfig{file=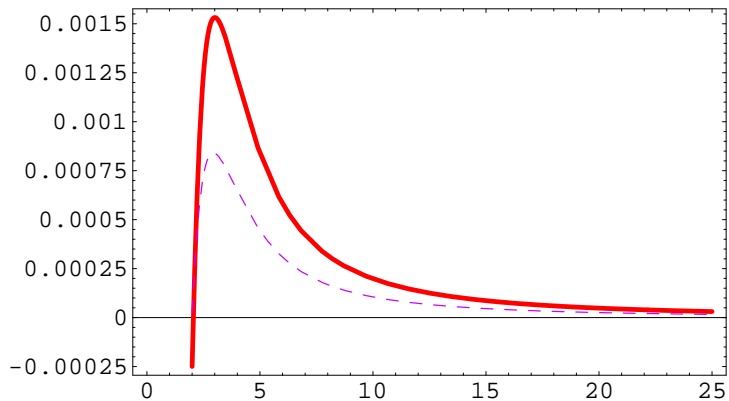,width=0.6\linewidth}
\caption{Plot of the difference $[m^2_{H_2}(\infty)-m^2_{H_2}]/4h_t^2N_c$
in units of $\ell$ as a function of $\ell\Lambda_s$, for
$\omega=1/2$ (solid) and $\omega=1/4$ (dashed).}
\label{fig}
\end{figure}

If we neglect the contribution (\ref{two}) we can write $m^2_{H_2}$ as:
\begin{eqnarray}
m_{H_2}^2&=&-\frac{h^2_t N_c}{8\pi^2\ell^{\,2}}
\sum_{n=-\infty}^{\infty}\left\{\pi^2(n+\omega)^2
\log\frac{(n+\omega)^2\pi^2+(\ell\Lambda_s)^2}{(n+\omega)^2\pi^2}\,\,
e^{-\frac{(n+\omega)^2\pi^2}{(\ell\Lambda_s)^2}}\right.\nonumber\\
&&\left. -\pi^2n^2
\log\frac{n^2\pi^2+(\ell\Lambda_s)^2}{n^2\pi^2}
e^{-\frac{n^2\pi^2}{(\ell\Lambda_s)^2}}\right\}.
\label{numeric}
\end{eqnarray}
Using again the Poisson resummation formula we can cast 
this expression into the form
\begin{equation}
m_{H_2}^2=-\frac{h^2_t N_c}{8\pi^3\ell^{\,2}}(\ell\Lambda_s)^3
\sum_{n=-\infty}^{\infty}
\left(e^{-2\pi in\omega}-1\right)\widetilde g(2n\ell\Lambda_s),
\label{three}
\end{equation}
where the function $\widetilde g(p)$ is the Fourier transform of 
\begin{equation}
g(y)=y^2e^{-y^2}\left[\log(1+y^2)-\log(y^2)\right].
\end{equation}
The dependence of $\widetilde g(p)$ for large Fourier modes
$p$ is obtained from the behaviour of the function $g(y)$ at the 
non-analytic point $y=0$.
We find that the leading term of $\widetilde{g}(p)$, in the limit
$|p|\to \infty$, is $-4\pi|p|^{-3}+\mathcal{O}(|p|^{-4})$. 
Thus, in the limit $\ell\Lambda_s\to \infty$, $m_{H_2}^2$ tends 
to~\footnote{Again the $n=0$ term is canceled by supersymmetry 
since $\widetilde g(0)$ is finite.}:
\begin{equation}
\label{analytic}
m_{H_2}^2(\infty)=\frac{h^2_t  N_c}{16\pi^2\ell^{\,2}}[
Li_3(e^{-2 i \pi \omega})+Li_3(e^{2 i \pi \omega})-2\zeta(3)]
+\mathcal{O}(\frac{1}{\ell\Lambda_s}),
\end{equation}
which agrees with the expression obtained using the 
KK-regularization~\cite{quiros}. Moreover,
the convergence of $m_{H_2}^2$ to $m_{H_2}^2(\infty)$ 
is very fast, as can be seen
from Fig.~\ref{fig}, where $\ell^{\,2}[m^2_{H_2}(\infty)-m^2_{H_2}]$
is plotted versus $\ell\Lambda_s$ for different values of $\omega$.

%%%%%%%%%%%%%%%%%%%%%%%%%%%%%%%%%%%%%%%%%%%%%%%%%%
%%%%%%%%%%%%%%%%%%%%%%%%%%%%%%%%%%%%%%%%%%%%%%%%%%

Even if the Gaussian distribution (\ref{gaussian}) we have assumed
for the localization of the brane along the extra dimension is a 
physical one, and strongly motivated by string theories, 
we would like to comment about the generality of our results 
with respect to
the distribution choice. Since the aim of a general distribution $f(y;l_s)$
is to regularize the $\delta$-function, a clear requirement the
function $f(y;l_s)$ must satisfy is that $\displaystyle \lim_{l_s\to 0}
f(y;l_s)=\delta (y)$. A simple example satisfying that requirement
is provided by the distribution
\begin{equation}
\label{counter}
f(y;l_s)=\frac{1}{\arctan(1/l_s)}\ \frac{l_s}{y^2+l_s^2}\ .
\end{equation}
Its Fourier transform over the orbifold length, $\tilde f(p;l_s)$,              
defines the $n$-dependent couplings in (\ref{l4})
as
\begin{equation}
\label{new-acoplos}
h_t^{(n)}=h_t\ \tilde f(\pi(n+\omega)/\ell;l_s)
\end{equation}
which again provides an exponential decoupling of heavy modes. 
In fact, for $l_s\ll \ell$ the function $\tilde f$ behaves as
\begin{equation}
\tilde f(p;l_s)\simeq\, 
e^{-|p|\, l_s}.
\label{anything}
\end{equation} 
while, unlike $e^{-|p|\, l_s}$, it is analytic at $p=0$.

Carrying out similar steps as in the Gaussian case leads to 
Eq.~(\ref{analytic}) with different subleading, $\mathcal{O}(1/\ell\Lambda_s)$,
corrections. We have found that analyticity of $\tilde f$ is an
essential ingredient for the UV insensitivity of the Higgs mass in a
supersymmetric theory. Moreover, any distribution with 
well defined moments, and satisfying the property that 
$\displaystyle\lim_{l_s\to 0} f(y;l_s)=\delta(y)$, 
should lead to a suppressed UV sensitivity for the one-loop Higgs mass.
We have checked this point by explicit calculations.

To conclude, we have proven that in the case of a Gaussian distribution the
sensitivity on the cutoff of the Higgs mass is suppressed at one loop
and no quadratic divergences appear in a supersymmetric theory. 
It is therefore fully justified
to exchange the (infinite) summation and the (infinite) integral as
done in the KK-regularization. Notice also that a similar calculation
can be done for the gauge interactions, and its contributions to the
Higgs mass, leading to a mode dependent gauge coupling
$
g^{(n)}=g \exp\{ -\frac{1}{2}\left(\frac{M^{(n)}}{\Lambda_s}
\right)^2\}
$ and a finite correction. Moreover radiative corrections to the mass of
other massless scalars (squarks, sleptons) localized on the brane also
lead to finite results. We have also shown that the same conclusions
hold for any well-defined distribution of the brane along the extra
dimension.

Let us finally notice that this result is also supported by explicit string
calculations~\cite{radiative}, where the squared Higgs mass is
obtained to be $\sim M_s^2$ (the string scale, $M_s$, playing the role of the
UV cutoff $\Lambda_s$) in the region $\ell M_s \simlt 1$ (stringy
region), and $\sim 1/\ell^{\,2}$ in the region $\ell M_s \gg 1$ (field theory
limit) and given by the expression (\ref{analytic}), in agreement with our
present results.

\section*{Acknowledgments} 
Two of us (AD and MQ) would like to thank I.~Antoniadis, K.~Benakli and
A.~Pomarol for many discussions on the subject.
The work of AD was supported 
by the Spanish Education Office (MEC) under an \emph{FPI} scholarship. 
The work of PJ was supported by the \emph{Deutsche Forschungsgemeinschaft}.
%%%%%%%%%%%%%%%%%%%%%%%%%%%%%%%%%%%%%%%%%%%%%%%%%%

\end{document}